\documentclass[pre,floats,superscriptaddress,showpacs,usenames]{revtex4}
\usepackage{amsmath}
\usepackage{graphicx}

\usepackage[dvipsnames]{color}

 \usepackage{array}
\usepackage{booktabs}
\usepackage{listings}
\usepackage{lmodern}
\usepackage{mathpazo}
\usepackage{microtype}

\usepackage{bm} 
\usepackage{amsmath}
\usepackage{amsbsy}

\newcommand{\be}{\begin{equation}} 
\newcommand{\ee}{\end{equation}}
\newcommand{\bea}{\begin{eqnarray}} 
\newcommand{\eea}{\end{eqnarray}}

\newcommand{\hbo}{H_G}
\newcommand{\hma}{H_C}
\newcommand{\sigmauno}{\left.\frac{d\hma}{dt}\right|_{M}}
\newcommand{\sigmadue}{\left.\frac{d\hma}{dt}\right|_{B}}

\begin{document}

 \title{About an H-theorem for systems with non-conservative interactions}

\author{Umberto Marini Bettolo Marconi}
\email{umberto.marinibettolo@unicam.it}
\address{ Scuola di Scienze e Tecnologie, 
Universit\`a di Camerino, Via Madonna delle Carceri, 62032 ,
Camerino, INFN Perugia, Italy}

\author{Andrea Puglisi}
\email{andrea.puglisi@roma1.infn.it}
\address{CNR-ISC and 
Dipartimento di Fisica, Universit\`a La Sapienza, p.le A. Moro 2, 00185 Rome, Italy}

\author{Angelo Vulpiani}
\email{angelo.vulpiani@roma1.infn.it}
\address{Dipartimento di Fisica, Universit\`a La Sapienza, and CNR - ISC, p.le A. Moro 2, 00185 Rome, Italy}

\begin{abstract}
We exhibit some arguments in favour of an H-theorem for a
generalization of the Boltzmann equation including non-conservative
interactions and a linear Fokker-Planck-like thermostatting term.
Such a non-linear equation describing the evolution of the single
particle probability $P_i(t)$ of being in state $i$ at time $t$, is a
suitable model for granular gases and is indicated here as
Boltzmann-Fokker-Planck (BFP) equation. The conjectured H-functional, which
appears to be non-increasing, is $H_C(t)=\sum_i P_i(t) \ln
P_i(t)/\Pi_i$ with $\Pi_i = \lim_{t \to \infty} P_i(t)$, in
analogy with the H-functional of Markov processes. The extension to
continuous states is straightforward.  A simple proof can be given for
the elastic BFP equation. A semi-analytical proof is also
offered for the BFP equation for so-called inelastic Maxwell
molecules. Other evidence is obtained by solving particular BFP cases
through numerical integration or through ``particle schemes'' such as
the Direct Simulation Monte Carlo.
\end{abstract}
\maketitle

\label{Introduction}
\section{Introduction}

The H-theorem is a consequence of the Boltzmann equation (whose
validity, at least for diluted gases, is now rather well understood)
and has a great relevance in statistical mechanics because it leads to
two basic results~\cite{B72}.  First it provides a dynamical proof of the
stationary Maxwell-Boltzmann distribution for the particle velocity
$p_{MB}({\bf v}) =[2\pi/(\beta m)]^{-3/2}e^{ -\beta m{\bf v}^2/2}$ and
second it states that the approach of the probability distribution
$p({\bf v},t)$ toward $p_{MB}({\bf v})$ is monotonic.  The latter
result can be interpreted as an instance of the 2-nd law of
the thermodynamics.

The literature about this subject is enormous and it is pretty
impossible to enter into details.  A discussion about
the deep and intriguing issues concerning the reversibility and
recurrence paradoxes, as well the rigorous derivation of the Boltzmann
equation in the Grad-Boltzmann limit, can be found in Refs.~\cite{C81,C06,CIP94,CFLV08} .

After the seminal results by Boltzmann obtained employing his
transport equation describing the behavior of diluted gases, similar H
theorems had been obtained for systems described by Markov processes,
that is systems ruled by master equations or Fokker-Planck equations~\cite{K61}.
There is however an important difference with respect to the
original Boltzmann H-theorem, where the functional $H_G[p]$ depends
only on $p({\bf v},t)$ and the knowledge of the asymptotic stationary probability
$p_{MB}({\bf v})$ is not required: for Markov processes the functional $H_C[p,\Pi]$ depends both on $p({\bf v},t)$
and the asymptotic stationary probability $\Pi({\bf v})$ which must be
determined, see next Section for details.
 
One of the basic feature of the Boltzmann equation is the presence of
bilinear terms, describing the binary collisions.  Although the
Boltzmann equation had been originally derived for systems with
conservative dynamics, it is not difficult, at least at formal level,
to write down a similar evolution equation for the one-particle
probability distribution for some dissipative systems (e.g. diluted
granular gases).  Due to the dissipative nature of granular gases, it
is necessary to introduce an external mechanism pumping energy into
the system, in order to have a statistical stationary state
(mathematically a non trivial stationary probability distribution) .
As a result the evolution equation for $p({\bf v}, t)$ includes a
linear term representing the coupling to the "external bath" and a
bilinear term accounting for binary collisions~\cite{PLMPV98}.
 
In the classical derivation of Boltzmann a key feature used to prove
the H-theorem is the presence of time reversal symmetry.  The absence
of such a property in dissipative systems is one of the technical
difficulties to derive a more general H-theorem.  Since the physical
importance and the mathematical relevance the study of relaxation
toward invariant probability, in terms of entropic functions, is an
interesting issue which attracted the interest of many
scientists~\cite{BCDVTW06,A04,T99,G05b}.

Previous studies showed that both the stationary and dynamical
statistical features are the combined results of bath and collisions~\cite{GSVP11b}.
We will see that in granular systems H-theorems, if any, must be the
outcomes of both the linear and bilinear part of the evolution
equation.  The control of such contributions is not easy: we are able
to show an H-theorem in the particular limit of "elastic granular"
gases in an external bath.  Such a system, although quite artificial,
presents nontrivial dynamical features.  In addition we give some
semi-analytical treatment of a Maxwell model of granular gas, and
detailed numerical computations supporting our idea.
  
The paper is organized as follows.  Section 2 is a quick summary of
known different H-theorems for the Boltzmann equation and Markov
processes.  In Section 3 we present a kinetic description of diluted
granular gases and some results about the H-theorems for such a
systems in particular limits.  Section 4 treats numerical simulation
of the granular gases in different regimes.  In Section 5 the reader
can find some conclusion. The Appendices are devoted to few technical
details.

\section{Monotonic approach to invariant probability}

For completeness we briefly review, in this Section, two known H-theorems. 

\subsection{Boltzmann equation for elastic isolated gases}

For the sake of notational simplicity we consider the case where the
states of the system are discrete.  Let call $P_i(t)$ the probability of
observing a state $i$ at time $t$.  Assuming the validity of the
molecular chaos assumption its evolution is governed by the following
non linear equation:
\be
{{d P_i(t)} \over {dt} }=\sum_{k,l,j} \Bigl[W^{(2)}_{(k,l)\to (i,j)} P_k(t) P_l(t)
-W^{(2)}_{(i,j)\to (k,l)} P_i(t) P_j(t) \Bigr],\label{uno}
\ee
where $W^{(2)}_{(k,l)\to (i,j)}$ denotes the transition rate of the
"collision" from the states $k$ and $l$ to the states $i$ and $j$.  By
invariance under time inversion one has the following property 
\be \label{trs}
W^{(2)}_{(k,l)\to (i,j)}=W^{(2)}_{(i,j)\to (k,l)} \ee 
so that \eqref{uno} takes the form 
\be \label{tre} {{d P_i(t)} \over {dt} }=\sum_{k,l,j} W^{(2)}_{(k,l)\to
  (i,j)} \Bigl[ P_k(t) P_l(t) - P_i(t) P_j(t) \Bigr] 
\ee

using such a structure it is easy to show the H-theorem, i.e.  the
$\hbo(t)$ function \be \hbo(t)=\sum_i P_i(t) \ln P_i(t) \ee is
monotonically decreasing:
\begin{equation}
 \label{boltz}
{ {d \hbo(t)} \over dt}\le 0
\end{equation}
and reaches its minimum $P_j=\Pi_j$ when the PdF corresponds to the
Maxwell-Boltzmann distribution.

\subsection{Markov processes}
\label{mapr}

An H-theorem also holds 
under rather general hypothesis~\cite{K61}  for processes governed by a Master equation 
of the form
\be \label{eq:master}
{{d P_i(t)} \over {dt} }=\sum_{k\ne i} \Bigl[W^{(1)}_{k\to i} P_k (t)
-W^{(1)}_{i\to k} P_i(t) \Bigr]            
\ee 
Here $W^{(1)}_{k \to i}$ is the transition rate from the state $k$ to the state $i$.

In this case indicating with $\{ \Pi_j \} $ the invariant probabilities , which are the solutions
of the following equation:
$$
\Pi_i={ 1 \over \gamma_i} 
\sum_{k\ne i} W^{(1)}_{k\to i} \Pi_k  \,\,\,\, where \,\,\,\,
\gamma_i= \sum_{k\ne i} W^{(1)}_{i\to k}  
$$
one has that the function $\hma(t)$
$$
\hma(t)=\sum_i P_i(t) \ln { P_i(t) \over \Pi_i}
$$
is non increasing, i.e.
\begin{equation} \label{mark}
{ {d \hma(t)} \over dt}\le 0
\end{equation}
and attains its minimum when $P_j=\Pi_j$ .  Let us note that $-\hma$
is the conditional entropy, also known as Kullback-Leibler or relative
entropy~\cite{CT06}. 

In the case of continuous variables it is sufficient to replace the
probability $P_i(t)$ with the probability density $p(x,t)$ (where $x$
is a state in a continuous space, e.g. velocity of a particle) and the
sums with the integrals, the $\hma$ function becomes
$$
\hma(t)=\int  p(x,t) \ln { p(x,t) \over \Pi(x)} dx \,\,.
$$

\subsection{Some remarks}

Let us notice that $\hma$ is an intrinsic property, that is switching
to a new variable $y(x)$, if the transformation between $x$ and $y$ is
invertible, $\hma$ is invariant, whereas $\hbo$ is not~\cite{CT06,MT11}.

Moreover, for certain classes of Markov processes it is possible to show~\cite{MT05} a result
stronger than the monotonic behavior of $\hma(t)$: there exists a
constant $\gamma >0$ such that 
\be \hma(t)\le e^{-\gamma t} \hma(0)
\,\, .  
\ee 

Finally, one may wonder if $\hma$ is non-increasing also in the case of the
(elastic) Boltzmann case, Eq.~\eqref{tre}. It is immediate to verify that it is true,
indeed:
$$
\hma(t)=\hbo(t)- \sum_i P_i(t)\ln\Pi_i
$$ and being $\ln \Pi_i$ a linear combination of conserved quantities,
the second term of the right hand side is constant, so that $d
\hma/dt=d \hbo/dt$.

In Appendix~\ref{app:proof} we report two derivations, necessary for
the proof of result~\eqref{eldecr} below. Such derivations are nothing
but the well known proofs of Eq.~\eqref{boltz} and Eq.~\eqref{mark}:
the reader may verify that, by replacing $f_i(t)$ with $P_i(t)$ and
$A_{i,j,k,l}$ with $W^{(2)}_{(i,j)\to(k,l)}$ in Appendix~\ref{app1b},
one obtains the proof of Boltzmann H-theorem, and, replacing
$\left.\frac{d \hma}{dt}\right|_{M}$ with $\frac{d \hma}{dt}$ in
Appendix~\ref{app1a}, the H-theorem for Markov processes is proved.

\section{Granular gases with homogeneous energy injection}

In the case of dilute granular gases interactions are dissipative
(e.g. inelastic hard-core collisions) and the property in
Eq.~\eqref{trs} does not hold. One of the consequences of energy
dissipation is that the Boltzmann equation~\cite{NE98} without any
external energy input has usually a trivial asymptotic state (e.g. the
velocities of all particles vanish). Among the many models of energy
injection~\cite{WM96,NETP99,GHS92}, experiments~\cite{GSVP11b,PGGSV12}
have shown the relevance of a mechanism where all particles are
coupled with a random energy reservoir~\cite{PLMPV98}: such a simple
mechanism well reproduces the effect of an interaction of all the
particles with rough vibrating boundaries of the container.

In this model, under the hypothesis of molecular chaos and in the
discrete representation introduced above, the probability $P_i(t)$
obeys the following ``Boltzmann-Fokker-Planck'' (BFP) equation:
\begin{equation} \label{eq:gran}
{{d P_i(t)} \over {dt} }=\sum_{k} \Bigl[W^{(1)}_{k\to i} P_k (t)
-W^{(1)}_{i\to k} P_i(t) \Bigr] +
\sum_{k,l,j} \Bigl[W^{(2)}_{(k,l)\to (i,j)} P_k(t) P_l(t)
-W^{(2)}_{(i,j)\to (k,l)} P_i(t) P_j(t) \Bigr].
\end{equation}
As anticipated, when collisions are inelastic the time inversion
symmetry, Eq.~\eqref{trs}, does not hold, i.e.  $W^{(2)}_{(k,l)\to
  (i,j)}\ne W^{(2)}_{(i,j)\to (k,l)}$. Notice that, in general, the
existence of a solution satisfying detailed balance with respect to
the Markov rates $W^{(1)}$ is not required to guarantee a stationary
state of Eq.~\eqref{eq:gran}. However in the literature detailed
balance with respect to $W^{(1)}$ has been often assumed~\cite{PLMPV98} and,
for this reason, it is customary to consider such an energy injection
mechanism equivalent to coupling the gas with a heat bath.

For continuous variables Eq.~\eqref{eq:gran} is replaced by an evolution equation for the density $p(x,t)$:
\be \label{eq:granc}
{\partial \over \partial t} p(x,t)={\cal  L}_{FP} p(x,t) +{\cal C}_B(p,p) 
\ee 
where ${\cal L}_{FP}$ is a linear Fokker-Planck operator and ${\cal
  C}_B(p,p)$ is a bilinear integral operator (the inelastic
Boltzmann-collision integral~\cite{NE98}). The first
operator describes the interaction of the system with the heat-bath
necessary to render the system stationary, while the second describes
the collisions between the particles: the combination of the two
operators produce a non trivial velocity distribution. Up to our
knowledge, for Eq.~\eqref{eq:gran} (or its counterpart with continuous
velocities, Eq.~\eqref{eq:granc}), no kind of ``H-theorem'' is
known. The failure of the usual ``H-theorem'', that for the $\hbo$
functional, has been verified in~\cite{BCDVTW06}.  

Notice that, assuming a
relaxation toward equilibrium, i.e. $P_i(t) \to \Pi_i$, then it is
easy to show that at large times $\hma$ is non-increasing.  In fact,
by writing $P_i(t)=\Pi_i + \delta P_i(t)$ with $\delta P_i$ small, one
has
\begin{equation} \label{smalldev}
\hma(t)=\sum_i \Bigr( \Pi_i + \delta P_i(t) \Bigl) \ln \Bigr( 1 + {\delta P_i(t)  \over \Pi_i}  \Bigl)\simeq \sum_i { \delta P_i(t)^2 \over \Pi_i}.
\end{equation}

We also mention that, for particular granular models, it is possible to
prove that Eq.~\eqref{eq:granc} in the elastic limit becomes
equivalent to a Fokker-Planck equation: in that case the H-theorem for the $\hma$ functional is obviously verified~\cite{PT06}.

\subsection{An elastic granular gas}
\label{sect:elgran}
Let us analyze a particular limit of the previous model where the collisions are elastic , so that
  $W^{(2)}_{(k,l)\to (i,j)}= W^{(2)}_{(i,j)\to (k,l)}$. The governing equation reads
\be
{{d P_i} \over {dt} }=\sum_{k} \Bigl[W^{(1)}_{k\to i} P_k 
-W^{(1)}_{i\to k} P_i \Bigr] +
\sum_{k,l,j}   W^{(2)}_{(k,l)\to (i,j)} \Bigl[ P_k P_l-
P_i P_j \Bigr]
\ee
and in addition we assume that the invariant probability $\{ \Pi_i \}$ is a stationary
solution of both the linear master equation and the non-linear
Boltzmann equation: i.e.  for the $\{ \Pi_k \}$ satisfying Eq.~\eqref{eq:master}, one has
\be \label{cons}
\Pi_i \Pi_j =\Pi_k\Pi_l
\ee
if the $(i,j) \to (k,l)$ collision is allowed.

It is important to realize that, even if the invariant probability is
somehow trivial, the dynamics is not, since it depends upon the
interplay between the bath and the collisions which may happen, for
instance, on different timescales. The example discussed in
Sec.~\ref{sec:disc}, see Fig.~\ref{fig:el_b}, well illustrates this
point, by showing the non-trivial dynamics of $\hbo$ which is non-monotonic.

In such a system one shows that
\be \label{eldecr}
{d \hma(t) \over dt} \le 0 .
\ee
In fact, we can write
\be
{d \hma(t) \over dt}=\sigmauno+\sigmadue
\ee
where
\be
\sigmauno=\sum_{i,k} \Bigl[W^{(1)}_{k\to i} P_k 
-W^{(1)}_{i\to k} P_i \Bigr] \left(\ln {P_i(t) \over \Pi_i}+1\right)
\ee
and
\be
\sigmadue=
\sum_{i,k,l,j}   W^{(2)}_{(k,l)\to (i,j)} \Bigl[ P_k P_l-
P_i P_j \Bigr] \ln{ P_i(t) \over \Pi_i}.
\ee
Since $\Pi_i \Pi_j=\Pi_k \Pi_l$ we can rewrite $\sigmadue$ as
\be
\sigmadue=
\sum_{i,k,l,j}   W^{(2)}_{(k,l)\to (i,j)}  \Pi_k \Pi_l  \Bigl[ { P_k P_l \over \Pi_k \Pi_l}-
{P_i P_j  \over \Pi_i \Pi_j  } \Bigr] \ln{ P_i(t) \over \Pi_i}.
\ee
It is now easy to show that both $\sigmauno$ and $\sigmadue$ are
negative: it is enough to follow the standard proofs of the H-theorems for the
Master equation and for the Boltzmann equation separately: those
proofs are reported in the Appendix A for completeness.

\subsection{Granular Maxwell model}

\label{sec:maxwell}

We discuss now an inelastic Maxwell model, a variation upon a theme,
originally proposed by Ulam~\cite{U80,E81} to study the approach to
equilibrium, introduced by Ben Naim and Kaprivski, as a minimal
kinetic model for granular gases~\cite{BK00,BCG00,BMP02} . The
advantage of this model is that all moments can be explicitly
computed~\cite{MP02}.  In the 1d thermostatted version of the model, Eq.~\eqref{eq:granc}, takes the form
\begin{equation}
 \partial_t p(v,t)\!=\Gamma\Bigl(\partial_v v p(v,t))
 +D(\partial^2_{v} p(v,t) \Bigr)+
 \frac{1}{\tau_c} \Bigl( \!\frac{2 }{1+\alpha}
 \!\int\!\!\!du\, p(u,t)p\left(\frac{2 v-(1-\alpha)u}{1+\alpha},t\right)-\!p(v,t) \!\Bigr)
\label{governing}
\end{equation}
where $\alpha \le 1$ is the restitution coefficient (when $\alpha=1$
the collisions are elastic), the first term on the right hand
describes the effect of the heat-bath at temperature $D/\Gamma$ and
corresponds to ${\cal L}_{FP}$, while the last term represents the non
linear collisional term, which is the sum of a gain term and a loss
term.

In general it is not possible to write explicitly $p(v,t)$, however,
it is possible to obtain the evolution law of its moments defined as
$\mu_{n}(t)=\int_{-\infty}^{\infty} dv v^n p(v,t)$ and prove that the
high velocity tails of the PdF are gaussian.  The evaluation of the
$H$-function requires the PdF, so that we used an approximation which
correctly reproduces all moments up to a given order and displays the
correct high velocity tails.  To achieve that, we consider the
following Sonine-Hermite representation of $p(v,t)$:
\begin{equation}
f(c,t)=\frac{1}{\sqrt{\pi}}e^{-c^2}[1+
\sum_{n=1}^{\infty}a_{n}(t) S_n(c^2)]
\label{fc}
\end{equation}
where $c^2= v^2/(2 \mu_2(t))$ is the non dimensional velocity squared
and $f(c,t)$ is the scaled PdF related to $p(v,t)$ by the
transformation
\be
p(v,t)=\frac{1}{\sqrt{2 \mu_2(t)}} \, f(c,t) \, .
\label{soninepdf}
\ee
For the sake of simplicity we assumed that the distribution is an even
function of the velocity so that all its odd moments vanish.  The
$S_n(c^2)$ are the Sonine polynomials of order $2n$, given by the
formula \cite{L03}:
$$
S_n(c^2)=\sum_{p=0}^n\frac{\Gamma(n+1/2)(-c^2)^p}{\Gamma(p+1/2)(n-p)!p!}
$$
having the property
\begin{equation}
\int_{-\infty}^{\infty} \frac{1}{\sqrt{\pi}}e^{-c^2} 
S_n(c^2)S_m(c^2)={\cal N}_n \delta_{m,n}=\frac{1}{\sqrt \pi}\frac{\Gamma(n+1/2)}{\Gamma(n+1)} \delta_{m,n}\label{ortho}
\end{equation}
and the $a_n(t)$ are coefficients of the expansion related to the moments $\mu_n(t)$ as shown in the appendix.
They can be calculated from  the averages
\begin{equation}
a_{n}(t)=\frac{1}{{\cal N}_n}\int_{-\infty}^\infty dc f(c,t) S_n(c^2) .
\label{an}
\end{equation}
Since the evaluation of the moments of a given order requires only the
knowledge of the moments of lower order one can proceed without
excessive difficulty to any desired order.  In practice, we carried on
our calculation up to the eighth moment. 

\begin{figure}
\includegraphics[width=10cm,clip=true]{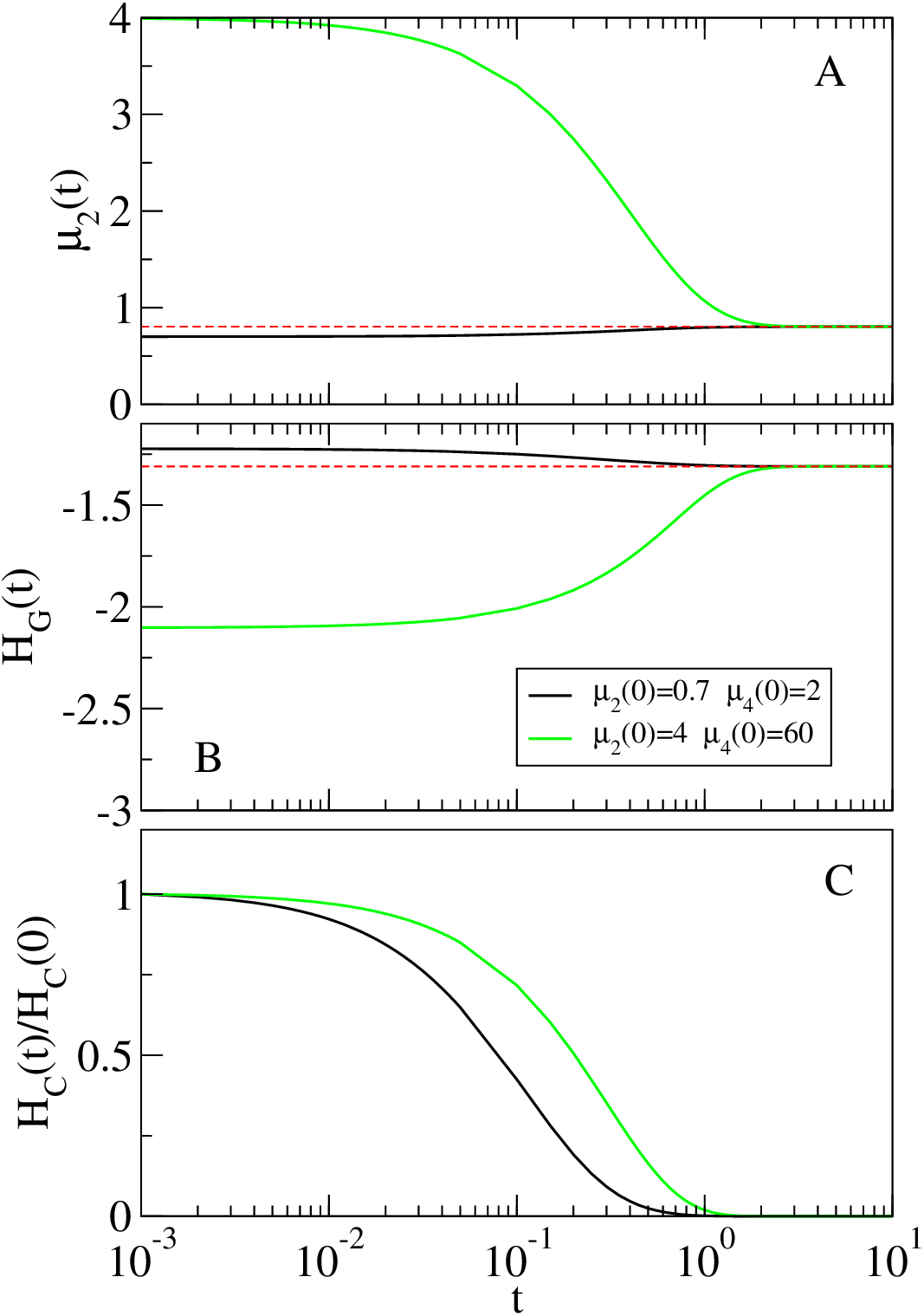}
\caption{\label{fig:umb}Inelastic Maxwell model with characteristic
  time $\tau_c=0.1$ combined with a thermal bath (characteristic time
  $\tau_b=1$ and temperature $T_b=1$) and $\alpha=0.95$, for two different initial conditions, where only $\mu_2(0)$ and $\mu_4(0)$ are not vanishing. A) Evolution
  of the average energy $\mu_2(t)$, B) Evolution of the Boltzmann
  $\hbo(t)$ function, C) Evolution of $\hma(t)$. }
\end{figure}

We can now evaluate the $H$-function for the Maxwell model using the
Sonine representation of $p(v,t)$ by performing numerically the
following integral:
\begin{equation}
\hma(t)=\int_{-\infty}^\infty dv p(v,t)\ln \left(\frac{p(v,t)}{\Pi(v)}\right) 
\label{hvp}
\end{equation}
where the asymptotic stationary distribution is obtained by inserting
in \eqref{soninepdf} the asymptotic values of the moments
$\mu_n(\infty)$, which are readily computed using
Eq.~\eqref{momentevolution} in Appendix~\ref{bettolo}:
\begin{equation}
\Pi(v)=\frac{1}{\sqrt{2\pi \mu_2(\infty)}} e^{-\frac{v^2}{2\mu_2(\infty)}}
\left[1+ \sum_{m=2}^{\infty} a_{m}(\infty) S_m\left(\frac{v^2}{2\mu_2(\infty)} \right)\right].
\label{asymptoticpdf}
\end{equation} 
Explicitly $\hma$ reads:
$$
\hma(t)= \int_{-\infty}^\infty dc \frac{1}{\sqrt{\pi}}e^{-c^2} \left[ 1+\sum_{m=2}^{\infty}a_{m}(t) S_m(c^2)\right]
\times
$$
$$
\left\{-\frac{1}{2}\ln\left(\frac{\mu_2(t)}{\mu_2(\infty)}\right) -c^2\left(1-\frac{\mu_2(t)}{\mu_2(\infty)}\right)+\ln\left[ 1+\sum_{l=2}^{\infty}a_{l}(t) S_l(c^2)\right] 
- 
\ln\left[1+\sum_{l=2}^{\infty}a_{l}(\infty) S_l\left(c^2 \frac{\mu_2(t)}{\mu_2(\infty)}\right)\right] \right\}.
$$ 
In Figs. \ref{fig:umb} we display the behavior of the $\hma$ and
$\hbo$ functions together with the evolution of the second moment of
the PdF for two different initial conditions but having the same
steady state. In the first case the second moment decreases towards
its asymptotic value, while the $\hma$ function also decreases,
whereas the $\hbo$ function increases.  In the second case instead the
$\mu_2(t)$ increases and both $\hbo$ and $\hma$ decrease in time.  The
two examples show that it is necessary to consider always $\hma$,
while $\hbo$ is not always monotonically decreasing.

A fast and simple way to prove numerically the hypothesis that the $\hma(t)$
function is always decreasing is to consider its evolution for a short
time interval $\Delta t$ starting from a distribution $p_{init}(v)=p(v,0)$
of the form \eqref{soninepdf} at the instant $t=0$ generated assuming
initial values of the moments arbitrarily with the only constraint
that this PdF is everywhere non negative.  We then compute the
evolution of the PdF over a small time interval, $\Delta t$, using the
governing equations for the moments $\mu_n(t)$ and finally calculate
the variation $\Delta \hma(t)=\hma(t+\Delta t)-\hma(t) $. For all
possible choices of the initial values we have found that such a
variation is which turns out to be always non positive.  It is worth
to comment that it is not necessary to follow the system evolution
over a longer time interval, to verify the persistence of the sign of
$\Delta \hma(t)$. In fact, any possible distribution $p(v,t^*)$ which
can be reached at time $t^*>0$ by the dynamics starting from the
distribution $p_{init}(v)$, represents itself a good candidate as
initial distribution, whose choice is arbitrary.  We have sampled a
large number of initial conditions $p_{init}(v)$ of the form
\eqref{soninepdf} randomly generated and verified that $\hma(t+\Delta
t)-\hma(t) \leq 0$.  

We conclude this section by saying that although we cannot prove analytically that for the
inelastic Maxwell there exist an H-theorem, our numerical results
provide a strong evidence that this is the case. The results of this
section should be compared with those obtained for other granular
systems, as discussed in the next Section.

\section{Numerical evidence for other examples of granular systems }

\subsection{Granular gases with discrete states}

\label{sec:disc}

We present here the simulation of a discrete BFP
equation~\eqref{eq:gran} for a choice of (conservative or
non-conservative) collisions and with the presence of a thermal
bath. In particular we have assumed that each state $i \in [0,M]$
represents a possible value of the single particle energy
$\epsilon_i$, for instance $\epsilon_i = i \delta E$ where $\delta E$
is some amount of energy. Collisions have been assumed to mix energy
between colliding particles and, optionally, to dissipate a part of
it, in order to reproduce the inelasticity in granular
gases. 

The collision model we adopted assigns the following collision rule to
transform colliding energies $i,j$ into post-collision energies
$i',j'$:
\begin{align}
i'=j+1-\Delta \\
j'=i-1-\Delta,
\end{align}
with $\Delta \ge 0$ being the amount of dissipated energy and the
additional condition $i' \ge 0,j' \ge 0$ is enforced. Just for simplicity we assume that the probability
of two particles of being chosen for a collision is independent of the
relative velocity, as it occurs in so-called Maxwell models previously discussed~\cite{E81,BCG00,BMP02}. The single particle mean free time between
collision is defined as $\tau_c$. The stated collision model determines
the rates $W^{(2)}_{(k,l)\to (i,j)}$ in the
equation~\eqref{eq:gran}. To avoid cumbersome expressions we do not
reproduce here such rates. 

The action of the heat bath is taken into
account through the {\em linear} part of Eq.~\eqref{eq:gran}. In
particular we have assumed that
\begin{equation}
W^{(1)}_{k\to i}=\frac{1}{\tau_b}e^{-\frac{1}{2T_b}(i-k)},
\end{equation}
where we have introduced $\tau_b$, the characteristic bath time, and
$T_b$, the bath temperature. Rates $W^{(1)}_{k\to i}$
satisfy detailed balance with respect to the ``equilibrium''
stationary distribution $\Pi^{eq}_i = c e^{-i/T_b}$ with $c$ a
normalization constant. Such equilibrium stationary distribution is
attained when collisions are switched off (i.e. $\tau_c \to \infty$)
as well as when they are elastic (i.e. $\Delta = 0$).

We have simulated Eq.~\eqref{eq:gran} with the rates discussed above
by means of a fourth order Runge-Kutta integration. 

\begin{figure}
\includegraphics[width=15cm,clip=true]{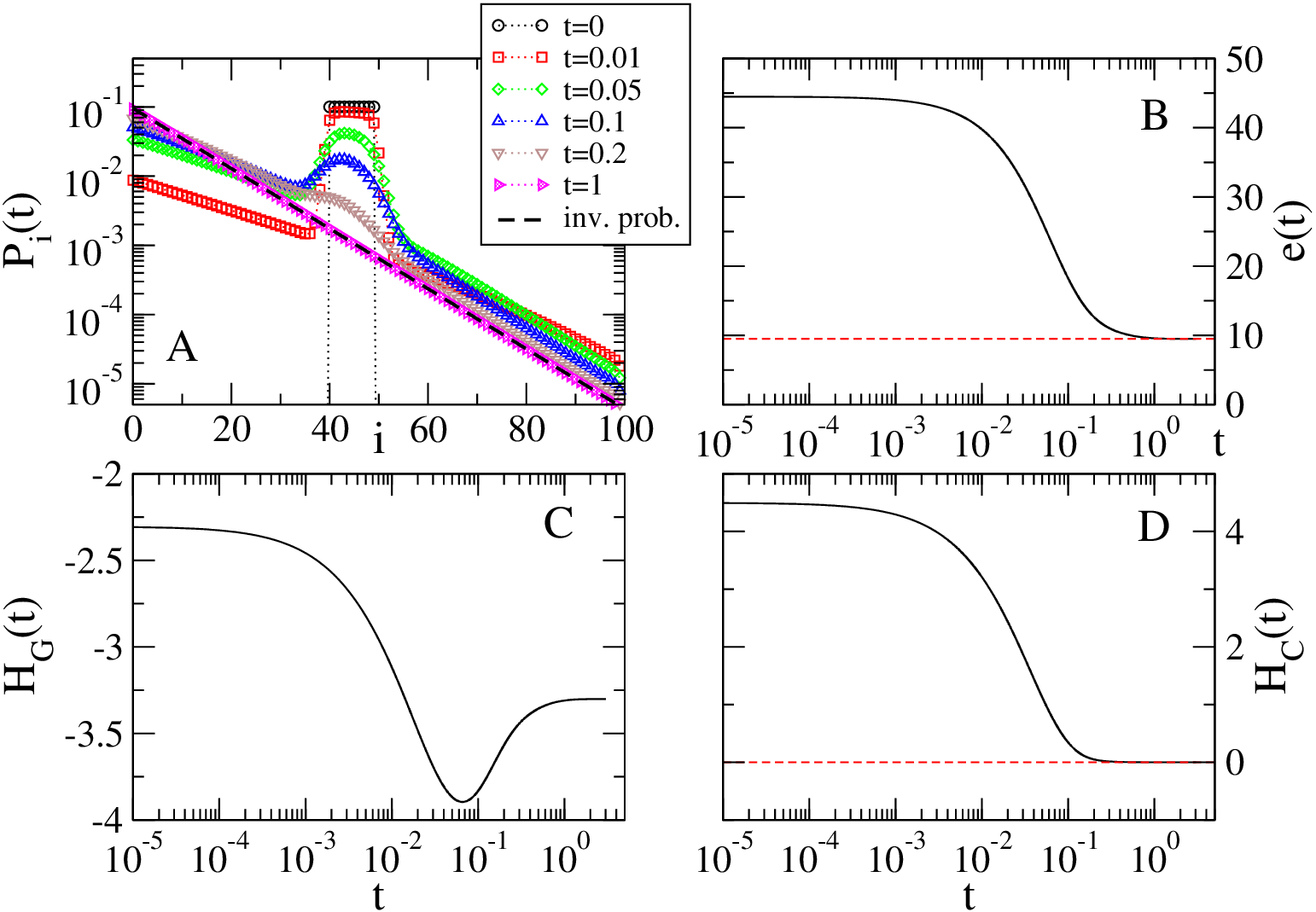}
\caption{\label{fig:el_b}Elastic collisions with characteristic time $\tau_c=1$ combined with a thermal bath (characteristic time $\tau_b=10$ and temperature $T_b=10.$). A) Evolution of $P_i(t)$, B) Evolution of the average energy $e=\sum i P_i$,  C) Evolution of the Boltzmann $\hbo$ function,  D) Evolution of the $\hma$ function}
\end{figure}

In Fig.~\ref{fig:el_b} we show that when the collision are elastic ($\Delta=0$) but
a thermal bath acts on each particle, one has a monotonic decrease of $\hma(t)$ but not of $\hbo(t)$. Note that
both mechanisms (elastic collisions and thermal bath), separately,
guarantee the existence of an equilibrium steady state: however the
thermal bath enforces a well defined temperature $T_b$, while
elastic collisions do not (the temperature, in the absence of the bath, would be chosen by initial
conditions, i.e. initial average energy). When the two mechanisms act
together, the final temperature is that of the bath, $T_b$. Panel
a) of Fig.~\ref{fig:el_b} shows the evolution of $P_i$ (from initial
conditions concentrated uniformly between $i=39$ and $i=49$) toward such an asymptotic equilibrium
distribution. Correspondingly, panel b) shows the evolution of the
average energy $e(t) = \sum_i i P_i$ of the system. Panel c) 
shows the {\em non-monotonic} behavior of the usual Boltzmann $\hbo$
function. Finally panel d) shows the fact, anticipated in
Section~\ref{sect:elgran}, that the evolution of $\hma$ is
non-increasing.

\begin{figure}
\includegraphics[width=15cm,clip=true]{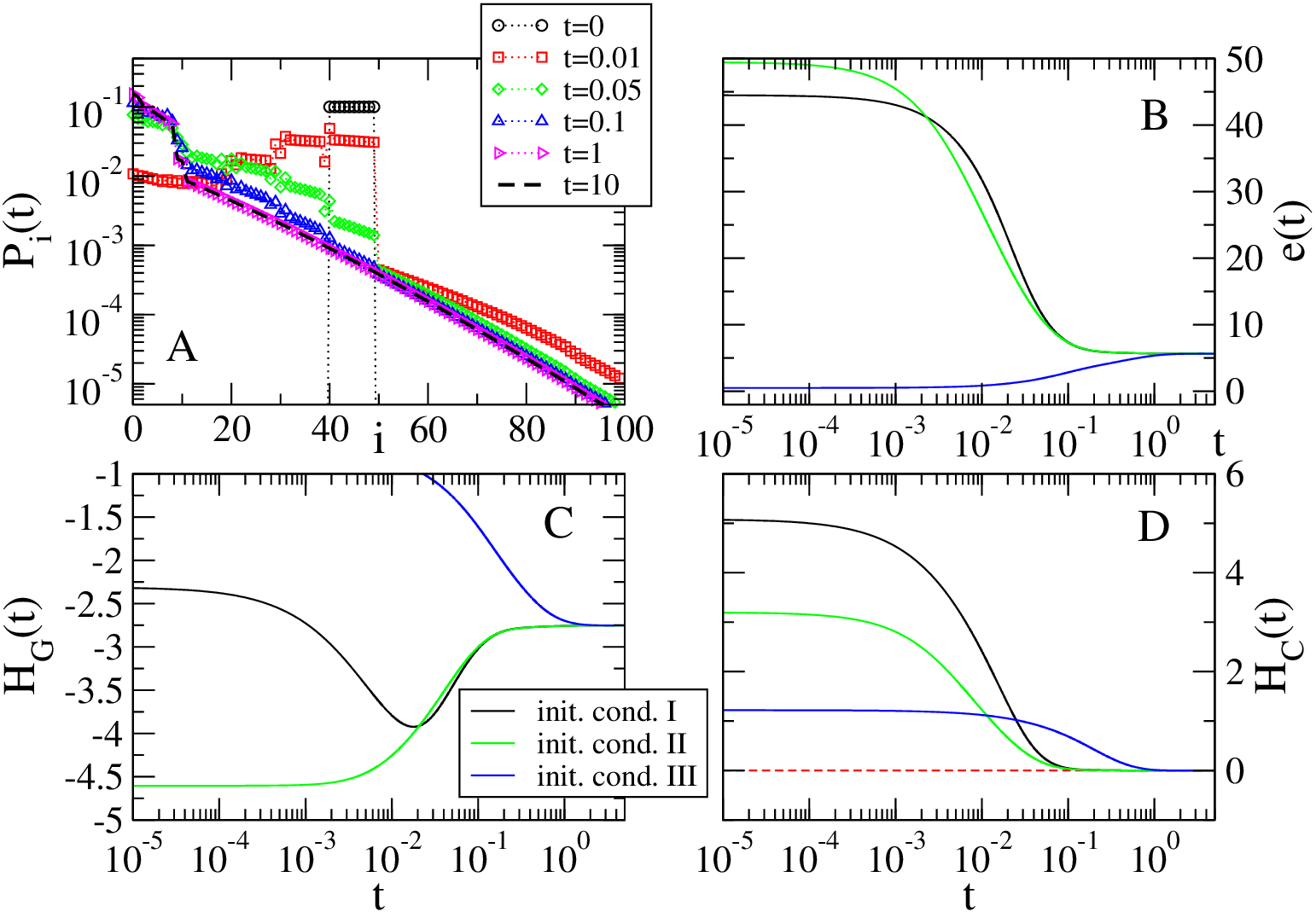}
\caption{\label{fig:inel2}Inelastic collisions ($\Delta=10$) with characteristic time $\tau_c=1$ combined with a thermal bath (characteristic time $\tau_b=10$ and temperature $T_b=10.$). Three different initial conditions are considered: i) $P_i(0)=1/11$ for $i\in[39,49]$ and $P_i(0)=0$ otherwise, ii)  $P_i(0)=1/100$ for $i\in[0,99]$ and $P_i(0)=0$ otherwise, iii) $P_i(0)=1/100$ for $i\in[0,99]$ and $P_i(0)=0$ otherwise. A) Evolution of $P_i(t)$ (shown only for initial condition i), B) Evolution of the average energy $e=\sum i P_i$,  C) Evolution of the Boltzmann $\hbo$ function,  D) Evolution of the $\hma$ function}
\end{figure}

In Fig.~\ref{fig:inel2} we report the results for the case with
dissipation ($\Delta>0$), in the presence of a thermal bath in order
to guarantee the attainment of a steady state. In this case the
knowledge of $\Pi_i$ is not known a priori, therefore a first long run
of the simulation is used to obtain it. A second run is then used to
measure $\hma$. We have repeated the simulation for three different
initial conditions, as detailed in the figure caption.  The evolution
of the probability distribution for the particular case starting with
$P_i(0)$ concentrated between $i=39$ and $i=49$ is shown in panel a)
of the Figure: it reaches an asymptotic distribution (the same for all
initial conditions) different from the equilibrium one.  In panel b)
we observe the average energy $e(t)$ which settles to a value smaller
than the bath temperature $T_b$ because of the inelasticity of
collisions. Panel c) and panel d) show that, while $\hbo$ not always
verifies the H-theorem, the validity of the H-theorem for $\hma(t)$ is
always verified. 

\subsection{Granular gases with continuous states through the Direct Simulation Monte Carlo}

Direct Simulation Monte Carlo (DSMC)~\cite{B94,CIP94} is
usually considered an effective ``solver'' for Boltzmann equations and
has been frequently used in the study of the kinetics of granular
gases. It is a so-called ``particles method'', since a finite number $N$ of
particles is evolved stochastically: the statistics of those $N$
particles approximates, as $N \to \infty$ the solution of the
corresponding Boltzmann equation. For the purpose of the present
paper, therefore, the study of the evolution of $\hma$ during the DSMC
dynamics, with non-conservative interactions and with the presence of
a heat bath, is meaningful as $N$ becomes larger and larger.

Here we use the DSMC algorithm discarding any spatial information,
however the algorithm is often used with spatial coordinates, by
dividing space in small cells. Our choice is equivalent to consider
the space-homogeneous version of the BFP equation~\eqref{eq:granc}. In
the algorithm time is advanced in time steps of length $\delta t$. At
each time step two sub-steps are performed: 1) the heat bath step,
where the velocity of each particle $i$ is advanced, from $t$ to
$t+\delta t$, by the discretized solution of an Ornstein-Uhlenbeck
stochastic process , i.e.  \be v_i(t+\delta t) = e^{-\frac{\delta
    t}{\tau_b}}v_i(t)+\sqrt{T_b\left(1-e^{-2\frac{\delta
      t}{\tau_b}}\right)}\phi_i(t), \ee where $\phi_i(t)$ is a random
variable extracted from a normal distribution (different $t$ and
different $i$ are all independent), $T_b$ is the bath temperature and
$\tau_b$ is the typical interaction time of a particle with the bath;
2) collisions are performed by choosing random couples $i,j$ of
particles and changing their velocities by the rule
\be \label{colrule} v_i' = v_i-\frac{1+\alpha}{2}(v_i-v_j),
\;\;\;\;\;\;\;v_j' = v_j+\frac{1+\alpha}{2}(v_i-v_j), \ee and the rate
of collision per particle is fixed at $1/\tau_c$. The parameter
$\alpha \le 1$ represents the restitution coefficient, when $\alpha=1$
collisions are elastic, otherwise they dissipate part of the kinetic
energy. Note that in this version of the DSMC the random choice of
particles is done uniformly: this is equivalent to solve a Boltzmann
equation with a collision probability {\em independent} from the
relative velocity of the colliding particle, as it happens in the
Maxwell models discussed in Section~\ref{sec:maxwell}.

\begin{figure}
\includegraphics[width=15cm,clip=true]{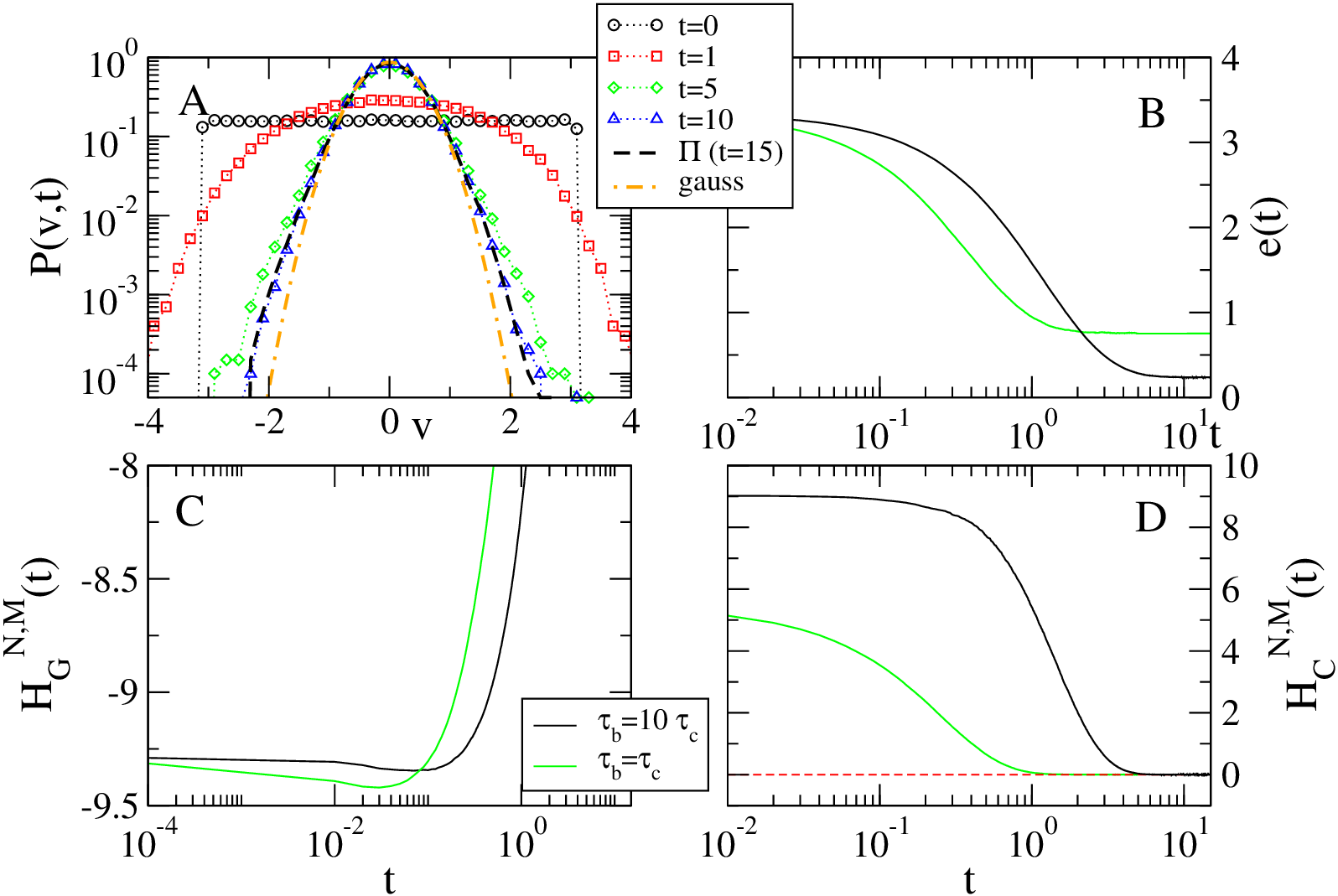}
\caption{\label{fig:dsmc}DSMC simulation of inelastic collisions
  ($\tau_c=1$ and $\alpha=0.6$) and thermal bath ($T_b=1$ and two
  different cases $\tau_b=1$ or $\tau_b=10$), with $N=10^5$ and
  $M=10^2$. A) Evolution of $P_i(t)$ in the case $\tau_b=10\tau_c$, B) Evolution of the average
  energy $e=\sum i P_i$, C) Evolution of the Boltzmann $\hbo$
  function, D) Evolution of the $\hma$ function}
\end{figure}

The results of the DSMC dynamics for a case with inelastic
with heat bath are shown in Fig.~\ref{fig:dsmc}. In panel a) we have
reported the evolution of the probability distribution, which is
started from a Gaussian distribution and $T_0 \gg T_b$. In panel b)
the evolution of the average energy is shown, demonstrating that the
initial temperature is forgot and the system attains a stationary
state with an average kinetic energy (also called granular
temperature) $T_g<T_b$, because of inelastic collisions. In frame c)
and d) we have reported the evolution of 
\begin{equation}
\hbo^{N,M}(t)=\sum_{i=1}^M P_i(t) \ln P_i(t)
\end{equation} 
and 
\begin{equation}
\hma^{N,M}(t)=\sum_{i=1}^M P_i(t) \ln \frac{P_i(t)}{\pi_i}
\end{equation}
where $P_i(t)$ and $\pi_i$ are the empirical velocity probability and
its long time limit respectively: the empirical probability at instant
$t$ is obtained by choosing a velocity interval $[-v_{max},v_{max}]$
with $v_{max}=10\sqrt{T_b}$ and dividing it in $M$ sub-intervals, and
taking $P_i(i)$ to be the number of particles with velocity in the
$i$-th sub-interval. For $N\gg 1$ and $M\gg 1$, we have $P_i(t) \simeq
p(-v_{max}+i\delta v,t)\delta v$ where $\delta v=\frac{2 v_{max}}{M}$,
so that 
\begin{align}
\int p(v,t) \ln p(v,t) dv &\simeq \hbo^{N,M}(t)-\ln(\delta v)\\
\int p(v,t) \ln \frac{p(v,t)}{\Pi(v)} &\simeq \hma^{N,M}(t).
\end{align}
The results reported in frame C and D of Fig.~\ref{fig:dsmc}
demonstrate the failure of the usual Boltzmann H-theorem (the one for
$\hbo^{N,M}$) together with the validity of $\frac{d}{dt}\hma^{N,M} \le 0$.

It is interesting to notice that in the examples shown in
Figures~\ref{fig:el_b},~\ref{fig:inel2} and~\ref{fig:dsmc}, the $\hbo$
functional has a behavior constituted by a first decrease followed by
an increase. This seems a consequence of the relatively fast action of
collisions superimposed to a slower action of the heat bath:
collisions (even if inelastic) do not change dramatically the energy
and therefore mainly contribute to ``equilibrate'' the initial
distribution, so that $\hbo$ satisfies the original H-theorem
($\frac{d\hbo}{dt} < 0$) at the beginning; when the heat bath action becomes
dominant, the distribution is near to the equilibrium one, i.e. $\hbo
\approx -\langle v^2 \rangle$ and the effect of the bath (a Ornstein-Uhlenbeck process
which, being linear, conserves the Gaussian shape), is mainly a
decrease of energy (initiated higher than $T_b$), implying $\frac{d\hbo}{dt} > 0$.

\subsection{The connection between the $\Gamma$ space and the $\mu$ space}

It is interesting to notice that the DSMC algorithm is a {\em Markov
  process} and therefore, as discussed in Section~\ref{mapr}, we
already know that an H-theorem holds. Nevertheless the DSMC is a
Markov process for an $N$-dimensional vector, i.e. in the so-called
``$\Gamma$ space'', while the BFP
Eq.~\eqref{eq:granc} governs the evolution of $p(v,t)$ i.e. the single
particle velocity distribution, i.e. it lives in the ``$\mu$
space''. 

By defining $\mathcal{P}(v_1,...,v_N,t)$ the probability density at time $t$ in the $N$-dimensional space, we know that for the DSMC it obeys a Master Equation of the kind
\be \label{fpbird}
\frac{\partial \mathcal{P}(v_1,...,v_N,t)}{\partial t}={\cal L}_N\mathcal{P}(v_1,...,v_N,t)
\ee
and it is customary to assume that it reaches a stationary state $\Pi(v_1,...,v_N)$.
Therefore the $\mathcal{H}_C$ function
\be
\mathcal{H}_C(t)=\int \mathcal{P}(v_1,...,v_N,t) \ln \frac{\mathcal{P}(v_1,...,v_N,t)}{\Pi(v_1,...,v_N)} dv_1...dv_N,
\ee
is a non-increasing function of time, i.e. $\frac{d\mathcal{H}_c}{dt}\le 0$.

Let us discuss the connection between the DSMC
and Eq.~\eqref{eq:granc} for our particular model. A more general and
rigorous proof can be found, for the elastic Boltzmann equation,
in~\cite{W91}.

In the case of the DSMC discussed above, which includes the
interaction with an external energy injection mechanism and pairwise
collisions, one may separate
\be \label{eq:gamma}
{\cal L}_N = \sum_{i=1}^N {\cal L}_{FP}(v_i) + \sum_{j=1}^N \sum_{i \neq j}^N T(v_i,v_j),
\ee
where ${\cal L}_{FP}$ is the operator representing the single particle Master equation, in our case that for the Ornstein-Uhlenbeck process, while 
\be
T(v_i,v_j)=\frac{1}{\alpha^2}b^{-1}(v_1,v_2)-1
\ee
is the operator for the collision between the $i,j$ particles, where
$b^{-1}(v_1,v_2)$ operates on a function of $v_1,v_2,...,v_N$ by mapping
$v_1,v_2$ into $v_1^*,v_2^*$ which are the pre-collisional velocities,
obtained by inverting Eq.~\eqref{colrule}. For non space-homogeneous
systems and hard-core interactions, the collisional operator is
different (for instance it enforces the condition of contact among
particles, as well as the velocity dependence of the scattering
probability, etc.), see for instance~\cite{B04}.

By marginalizing Eq.~\eqref{fpbird} and assuming Molecular Chaos,
i.e. $p_2(v_1,v_2,t)=p(v_1,t)p(v_2,t)$ for pre-collisional velocities,
it is simple to get the BFP Eq.~\eqref{eq:granc}.

It is interesting to notice that, by assuming a {\em stronger}
version of Molecular Chaos, valid for all velocities (i.e. not only the pre-collisional ones)
\be
{\cal P}(v_1,...,v_N,t)=\Pi_{i=1}^N P(v_i,t),
\ee
it is immediate to get
\be
{\cal H}_c(t) = N \int dv P(v,t) \ln \frac{P(v,t)}{\Pi(v,t)},
\ee
which implies 
\be
\frac{d \hma}{dt} \le 0.
\ee
We stress however that such a  {\em stronger} form of Molecular Chaos
is usually violated in non spatially
homogeneous granular gases, as shown in recent
experiments~\cite{GSVP11b,PGGSV12}. The fact that it could work in the
models discussed above is perhaps a consequence of spatial homogeneity
together with the action of the external heat bath.

\section{Conclusions}

In the present paper we have offered several examples of
Boltzmann-Fokker-Planck (BFP) models with conservative and non-conservative
interactions, where an H-functional of the kind
\begin{equation}
H_C(t)=\sum P_i(t) \ln \frac{P_i(t)}{\Pi_i}
\end{equation}
appears to be non-increasing for the whole evolution from arbitrary
initial conditions toward the asymptotic steady state $\Pi_i$.  The
only case where we are able to prove such a conjecture is the elastic
BFP case, where the proof is a ``superposition'' of the proofs of the
two different H-theorems for the elastic Boltzmann equation and for
Markov processes: notwithstanding the simplicity of the proof and the
triviality of the steady state, the elastic BFP model has a
non-trivial dynamics where collisions and thermostat may act on
different timescales, as demonstrated by the non-monotonous behavior
of the Boltzmann $H_G(t)$ functional.  Establishing the
monotonicity of $H_c(t)$ or other entropic functionals for a rather
general class of models is certainly a challenge for future
research~\cite{A04,V06b}. In a remark concluding the last section, we
have recalled that the BFP equation may be obtained by marginalizing a
master equation for the Markovian evolution of the many-particles
vector in $\Gamma$ space: this remark, which is behind the operating
principle of DSMC schemes, could be a possible starting point for
further investigations on this complex and fascinating subject.

\begin{acknowledgments}
The authors acknowledge useful discussions with A. Baldassarri. They also acknowledge the support of the
Italian MIUR under the PRIN 2009 grant n. 2009PYYZM5. A. P. acknowledges the support of the
Italian MIUR under the grant FIRB-IDEAS n. RBID08Z9JE.
\end{acknowledgments}

\appendix

\section{H-theorem in the case of conservative interactions and under the action of a heat bath}
\label{app:proof}

\subsection{Master equation contribution}
\label{app1a}

Let us introduce the variables $f_i(t)=P_i(t)/\Pi_i$ and $B_{ik}=W^{(1)}_{k\to i}\Pi_k$. It is also useful to
define the function $F(x)=x \ln x$, so that $F'(x)=1+\ln x$. It is easy to verify that 
\begin{equation}
\sigmauno=\sum_{i,k} F'(f_i)(B_{ik}f_k-B_{ki}f_i)=\sum_{i,k}B_{ik}[f_kF'(f_i)-f_kF'(f_k)], \label{stat}
\end{equation}
where we have exchanged the indexes in the second part of the sum, in order to collect $B_{ik}$. Then one notices that for a set of numbers $\psi_i$ the following relation holds
\be \label{zero}
\sum_{i,k}B_{ik}(\psi_i-\psi_k)=0,
\ee
which is a consequence of stationarity for $\Pi_i$. Adding Eq.~\eqref{zero} to the last line of Eq.~\eqref{stat}, and choosing $\psi_n=F(f_n)-f_nF'(f_n)$, one gets
\be
\sigmauno = \sum_{i,k}B_{ik}[(f_k-f_i)F'(f_i)+F(f_i)-F(f_k)].
\ee 
Now, since $F''(x)>0$, it is immediate to see that 
\be
[(f_k-f_i)F'(f_i)+F(f_i)-F(f_k)] \le 0,
\ee
and therefore $\sigmauno\le 0$. 
The proof shown here is that found in~\cite{K07}.

\subsection{Boltzmann equation contribution}
\label{app1b}
Here it is useful to introduce the variable $A_{i,j,k,l}=W^{(2)}_{(i,j)\to(k,l)}\Pi_i\Pi_j$.
The symmetry between collisions $(k,l)\to (i,j)$ and $(k,l) \to (j,i)$ implies the following identities $A_{k,l,i,j}=A_{i,j,k,l}=A_{l,k,j,i}=A_{l,k,i,j}$, so that
$$
\sigmadue={1 \over 2}
\sum _{i,j,k,l} A_{i,j,k,l}
\{ f_k(t)f_l(t)-f_i(t)f_j(t) \} \ln [f_i(t) f_j(t)] 
\,\, ,
$$
and by means of the symmetry between $(k,l)\to (i,j)$ and $(i,j) \to (k,l)$  one gets
$$
\sigmadue=-{1 \over 4}
\sum _{i,j,k,l} A_{i,j,k,l} 
\{ f_i(t)f_j(t)-f_k(t)f_l(t) \} 
\{ \ln [ f_i(t)f_j(t)]  -  \ln [f_k(t) f_l(t)] \} 
\,\, .
$$ 
We notice that $A_{i,j,k,l}>0$. Furthermore, since $ (\ln x - \ln
y)(x-y) \ge 0$ for every $x>0$ and $y>0$, we finally get $\sigmadue
\le 0$. The equality holds only when the $\{ f_i \}$ are identically
$1$, i.e. $P_i=\Pi_i$. The above proof is the standard one contained
in any textbook and is due to Ludwig Boltzmann~\cite{B72}.

\section{Time evolution of the moments for the Maxwell model}

\label{bettolo}
We provide here some details regarding the calculations presented in the main text.
 Applying the Fourier transform to eq. \eqref{governing}, 
 we obtain the following governing equation for characteristic function
 $\hat P(k,t)=\int_{-\infty}^{\infty} dv e^{ikv} p(v,t)$:

\begin{equation} \label{fou}
\partial_t \hat P(k,t) =
-D k^2\hat P(k,t) -\Gamma k \partial_k\hat P(k,t)
-\frac{1}{\tau_c}[\hat P(k,t)- \hat P(\gamma k,t) 
\hat P((1-\gamma k,t) ]
\end{equation}
with $\gamma=(1+\alpha)/2$.
Substituting the representation
$\hat P(k,t)=\sum_{n=0}^{\infty} \frac{(i k)^n}{n !} 
\mu_{n}(t)$ in \eqref{fou} and equating the coefficients of the same power of $k$
we derive a set of ordinary differential equations describing the
evolution of the moments, whose integration is straightforward and leads to expressions
of the type:
\begin{equation}
\mu_n(t)=e^{-K_n t}\mu_n(0)+e^{-K_n t}\int_0^t dt' R_n(t')\, e^{K_n t'}
\label{momentevolution}
\end{equation}
with
\be
K_n=n \Gamma - \frac{d_n}{\tau_c} 
\ee
where the coefficients $d_n$ are 
\begin{equation}
d_n =-1+[\gamma^n+(1-\gamma)^n]
\end{equation}
and 
$$
R_2=2 D, \;\;\;\;\; R_4(t)=A\mu_{2}(t)^2 +12 D \,\mu_2(t)
$$
$$
R_6(t)=B\mu_{2}(t)\mu_{4}(t) +30 D, \,\mu_4(t) \;\;\;\;\;\; R_8(t)=C_1\mu_{2}(t)\mu_{6}(t) +C_2\mu_{4}(t)^2 +56 D \mu_6(t)
$$
where the coefficients have the following expressions: 
$$A=6  \frac{[\gamma(1-\gamma)]^2}{\tau_c} \, , \;\;\;\;\; B=15  \frac{\gamma^2(1-\gamma)^2 [\gamma^2+(1-\gamma)^2]}{\tau_c}  $$
and 
$$C_1=28 \frac{ \gamma^2(1-\gamma)^2 [\gamma^4+(1-\gamma)^4]}{\tau_c}, \;\;\;\;\;\;\;C_2=70  \frac{\gamma^4(1-\gamma)^4}{\tau_c}  $$
Notice that $A,B,C_1,C_2,d_n$ vanish in the elastic limit $\gamma\to 1$
as the elastic collisions do not affect the PdF nor its moments.
Since in eq. \eqref{momentevolution} the evolution of the moment of order n is coupled only 
to the evolution of moments
of order smaller than n, the solution is very simple and can be achieved recursively.

In order to construct the 
PdF  we have to use the Sonine-Hermite expansion and compute the coefficients
$a_n(t)$ which are proportional to the cumulants of the distribution.
In terms of the moments we find:
\begin{subequations}
\begin{align} 
a_2(t)&=[-1+\frac{1}{3}\frac{ \mu_4(t)}{\mu_2(t)^2}] \label{aa2} \\
a_3(t)&=[-2+ \frac{\mu_4(t)}{\mu_2^2(t)}  - \frac{1}{15}\frac{\mu_6(t)}{\mu_2^3(t)}] \label{aa3}\\
a_4(t)&=[-3+ 2\frac{\mu_4(t)}{\mu_2^2(t)} - 
\frac{4}{15}\frac{\mu_6(t)}{\mu_2^3(t)} +\frac{1}{105}\frac{\mu_8(t)}{\mu_2^4(t)} ] \label{aa4} 
\end{align}
\end{subequations}
Notice that $a_2,a_3,a_4$ being proportional to the cumulants vanish for the Gaussian distribution,
but not for the granular gas.

\bibliography{fluct.bib}

\end{document}